\documentclass[prl,twocolumn,amsmath,amssymb,floatfix,superscriptaddress]{revtex4}
\usepackage{graphicx}
\usepackage{amsmath}
\usepackage{bm}

\newcommand{\bq}{\begin{equation}}
\newcommand{\ee}{\end{equation}}
\newcommand{\fr}[2]{\frac{#1}{#2}}
\newcommand{\eps}{\varepsilon}

\renewcommand{\vec}[1]{\mathbf{#1}}
\usepackage{colordvi}
\usepackage{graphicx}
\usepackage{color}

\begin{document}
\title{Magnetization current and anomalous Hall effect for massive
 Dirac electrons}
\date{\today }

\author
{P. G. Silvestrov} \affiliation{Institute for Mathematical
Physics, TU Braunschweig, 38106 Braunschweig, Germany}

\author{P. Recher}
\affiliation{Institute for Mathematical Physics, TU Braunschweig,
38106 Braunschweig, Germany} \affiliation{Laboratory for Emerging
Nanometrology Braunschweig, 38106 Braunschweig, Germany}


\begin{abstract}

Existing investigations of the anomalous Hall effect ({\it i.e.} a
current flowing transverse to the electric field in the absence of
an external magnetic field) are mostly concerned with the
transport current. However, for many applications one needs to
know the total current, including its pure magnetization part. In
this paper, we employ the two-dimensional massive Dirac equation
to find the exact current flowing along a potential step of
arbitrary shape. The current is universal, {\it i.e.} it depends
only on the asymptotic value of the potential drop. For a
spatially slowly varying potential we find the current density
$\mathbf{j}(\vec r)$ and the energy distribution of the current
density $\vec j^\varepsilon(\vec r)$. The latter turns out to be
unexpectedly nonuniform, behaving like a $\delta$-function at the
border of the classically accessible area at energy~$\eps$.
Consequently, even in a weak electric field the transverse current
density can not be described semiclassically. To demonstrate
explicitly the difference between the magnetization and transport
currents we consider the transverse shift of an electron ray in an
electric field.

\end{abstract}
\maketitle


\section{I. Introduction}

Currents flowing in topological insulators (TIs) are usually
associated with the electron gapless edge or surface
modes~\cite{KaneMele05,Bernevig06}. Other types of currents
present in materials with the nontrivial band structure, which now
are accessible
experimentally~\cite{GaimSci14,McEuenSci14,TaruchaNatPhys15}, stem
from the anomalous Hall effect
(AHE)~\cite{MacDonaldRMP10,NiuRMP10}. The standard approach to
describe the AHE is to employ the equation of motion for a wave
packet in a crystal~\cite{Chang1996PRB,Sundaram1999PRB}
 \bq\label{Berry}
\dot{\vec r} ={\partial \eps(\vec p)}/{\partial\vec p} -\vec v_B,
 \ee
with the Berry velocity $\vec v_B =({e}/{\hbar})\,\vec E\times
\vec\Omega(\vec p)$ normal to the local electric field and the
Berry curvature $\vec\Omega(\vec p)= i\hbar^2\vec\nabla_{\vec
p}\times\langle u_{\vec p}|\nabla_{\vec p}u_{\vec p}\rangle$
accounting for the change of the multicomponent wave function upon
moving in the Brillouin zone. What is frequently not appreciated,
even in the situations when Eq.~(\ref{Berry}) is applicable the
total microscopic current has two components named transport and
magnetization - associated solely with the wave packet rotation -
currents~\cite{CooperPRB97, XiaoPRL06}.
The Berry velocity is responsible only for the transport current
of electrons. Which current will be measured depends on the
particular experiment. In the case of nonequilibrium electron ray
injection,
the transport current described by Eq.~(\ref{Berry}) is observed.
In this paper, we consider the total equilibrium
current density, whose distribution can not be described by
Eq.~(\ref{Berry}), but which is responsible for the magnetic
moment of the electron gas and the interaction of electrons with
electro-magnetic fields. The microscopic current created in
response to an electric field generates a magnetic field, leading
to the Faraday effect~\cite{VolkovJETPL85, FaradayExp} and other
topological magneto-electric effects~\cite {Qi2008PRB}.

Rather surprisingly, we find that not only the magnetization
current for individual electrons may be large compared to the
transport one, but furthermore the different contributions to the
total current density of many electrons have a tendency to cancel
out. The only contribution to the microscopic current originate
from the electrons at the turning points (stopping points) where a
semiclassical description in terms of the wave packet dynamics is
not applicable (see Eq.~(\ref{distrib_current}) of our paper).

Specifically, we consider the two-dimensional massive Dirac
Hamiltonian, a paradigmatic model exhibiting the AHE found in
various material systems of current interest. In
time-reversal-symmetric
systems~\cite{NiuPRL07,GaimSci14,XiaoPRL12,McEuenSci14} there are
several Dirac cones and the anomalous bulk current is of the
valley-Hall type~\cite{LenskyPRL15}. A single massive Dirac cone
is realized on the surface of a three-dimensional TI covered by a
ferromagnetic insulator film~\cite{Qi2008PRB}. There, the AHE
corresponds to the charge Hall current, giving rise to the
fascinating physics of axion electrodynamics~\cite {Qi2008PRB}.

\begin{figure}
\includegraphics[width=7.5cm]{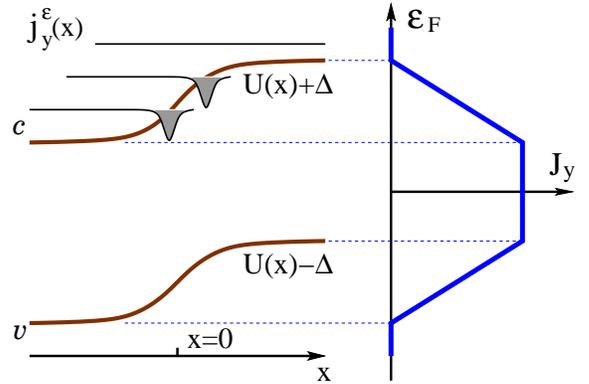}
\caption{\label{fig:1} Total dissipationless equilibrium anomalous
Hall current flowing in $y$-direction, $J_y$, along the potential
step $U(x)$ as a function of Fermi energy $\eps_{\rm F}$ for a
two-dimensional massive Dirac Hamiltonian with gap $2\Delta$. The
labels {\it v,c} denote valence- and conduction bands,
respectively. The energy distribution of the current density
$j^\eps_y(x)$ is also shown schematically for several values of
$\eps$.
 }
\vspace{-.3cm}
\end{figure}

We start in Sec.~II with the calculation of the total equilibrium
current flowing in $y$-direction along a potential step $U(x)$,
see Fig.~\ref{fig:1}. Provided that the potential is constant away
from the step, this calculation is exact and valid for any shape
of $U(x)$ and any value of the Fermi energy $\eps_{\rm F}$.

Our main findings, summarized in Eq.~(\ref{distrib_current}) of
section~III, concern the total AHE current density in slowly
varying electrostatic potentials.
For a smooth electrostatic potential $U(\vec r)$, we calculate the
current density $\vec j(\vec r)$ and the energy distribution of
the current density $\vec j^{\eps}(\vec r)$. Electron trajectories
with a given energy $\varepsilon$ in two-dimensions cover the
areas bounded by the lines of the classical stopping points
(having vanishing velocity, $\vec v=0$). We argue that the energy
distribution of the current density $\vec j^{\eps}(\vec r)$ has
the form of a $\delta$-function existing only along these lines of
stopping points ({\it cf.} Eq.~(\ref{distrib_current})). The
"quantum width" of this $\delta$-function scales like $\Delta
r\sim (\hbar^2/|\nabla U|)^{1/3}$, being nonperturbative in both
Planck's constant and the electric field.

At the end of the paper, in Sec.~IV, we present a semiclassical
calculation of the side jump of an electron ray traversing the
region with potential step $U(x)$ caused by the AHE, in agreement
with Eq.~(\ref{Berry}). The approach adopted in this section
allows us to distinguish and describe within the same calculation
both the magnetization part of the current and the pure transport
current. Although the content of section~IV may appear somewhat
methodological, we are not aware of other existing derivations of
the anomalous velocity Eq.~(\ref{Berry}), relying only on the
stationary solutions of the Schr\"{o}dinger(Dirac) equation,
without explicit consideration of the time-dependent wave-packet
evolution. Interestingly, by considering the stationary ray
dynamics (instead of deriving the wave packet equations of motion
Eq.~(\ref{Berry})~\cite{MacDonaldRMP10,NiuRMP10}) we were able to
find the entire electron's trajectory in a pedagogically appealing
way including the transverse (AHE) shift and show that the
magnitude of this shift has an upper bound.

\section{II. Microscopic calculation of the equilibrium
current}

We consider the Hamiltonian
 \bq\label{HamDirac}
H=v_{\rm F}(\sigma_x p_x +\sigma_y p_y) +\Delta\sigma_z +U(x) \ .
 \ee
Here, the Pauli matrices act on the sublattice, valley, or spin
degree of freedom depending on the material under consideration.
Instead of an uniform electric field, we consider a continuous
step-like potential $U(x)$, with $U_R=U(x\rightarrow\infty )$ and
$U_L=U(x\rightarrow-\infty )$. An arbitrarily large electric field
$e\vec E=-\vec\nabla U$ exists only in a certain region
around~$x\approx 0$. At large $|x|$, the potential is constant and
there is a gap in the spectrum $|\eps-U_{R,L}|<\Delta$, shifted up
on the right and down on the left side of the potential step, see
Fig.~\ref{fig:1}. For definiteness, we choose $U_R -U_L< 2\Delta$.
We rely here neither on semiclassical nor on weak or constant
electric field approximations.

Let $\Psi_j(x,y)= e^{ip_y y/\hbar}\psi_j(x)$ be a complete set of
eigenfunctions of the Hamiltonian Eq.~(\ref{HamDirac}) with
conserved momentum $p_{y_j}\equiv p_y$. Our goal is to find the
transverse current density defined via the velocity operator
$\dot{\vec r} =i[H,\vec r]/\hbar$ for the Dirac Hamiltonian with
$e=-|e|$
 \bq\label{current_def}
j_y(x)=e v_{\rm F}\langle \sigma_y\rangle = e v_{\rm
F}\sum_{i,\eps_i <\eps_{\rm F}} \psi_i^\dagger (x)\sigma_y
\psi_i(x) \ .
 \ee
Our approach to find $j_y(x)$ is motivated by the calculation of
the out of plane current induced polarization in Rashba wires in
Refs.~\cite{SilvestrovPRL09,SilvestrovImryBook10}. Consider the
spin-density for a particular stationary state $\psi^\dagger
(x)\sigma_x \psi(x)$,
 \bq
0\equiv \fr{d}{dt}\psi^\dagger \sigma_x \psi =
\fr{i}{\hbar}\psi^\dagger [ H,\sigma_x]\psi {-}v_{\rm F}\partial_x
(\psi^\dagger\psi) \ .
 \ee
The left and right sides of this equation follow from the
calculation of the time derivative in case of the time evolution
of the eigenfunction of Eq.~(\ref{HamDirac}) taken in two forms
$\psi(x,t)=e^{-iHt/\hbar}\psi(x)=e^{-i\eps t/\hbar}\psi(x)$. Then
we readily find
 \bq\label{spincurrent}
\Delta\,\psi^\dagger\sigma_y\psi - v_{\rm F}
p_y\psi^\dagger\sigma_z\psi = {-}\fr{\hbar v_{\rm F}}{2}\partial_x
(\psi^\dagger\psi) \ .
 \ee
This relation is enough to find the AHE transverse current for
electrons with $p_y=0$. The general case with $p_y\neq 0$ requires
more effort.

Since the conserved momentum $p_y$ and $\Delta$ enter the
Hamiltonian Eq.~(\ref{HamDirac}) in a similar fashion, it is
natural to consider a two parameter family of Hamiltonians
$H=H(\Delta,p_y)$. The two eigenfunctions $\psi(\Delta, \pm p_y)$
of two Hamiltonians $H(\Delta,\pm p_y)$ (still depending on the
coordinate $x$) are related to the eigenfunctions of the same
Hamiltonian with $p_y=0$ and enlarged mass
 \begin{align}\label{finitepy}
\psi(\Delta, \pm p_y) = e^{\pm
\fr{i}{2}\theta\sigma_x}\psi(\sqrt{\Delta^2+ v_{\rm F}^2p_y^2},0)
\ ,
 \end{align}
where $\tan\theta=v_{\rm F}p_y/\Delta$. Calculating the
expectation values of Eq.~(\ref{HamDirac}), we find the following
identity relating the expectation values for two solutions with
opposite $p_y$
 \begin{align}\label{rotatedspin}
& \Delta\psi^\dagger_+\sigma_z\psi_+ +v_{\rm F} p_y
\psi^\dagger_+\sigma_y\psi_+
\nonumber\\
=\,\, &\Delta\psi^\dagger_-\sigma_z\psi_- -v_{\rm F} p_y
\psi^\dagger_-\sigma_y\psi_- \ ,
 \end{align}
where $\psi_\pm =\psi(\Delta, \pm p_y)$. Here, we used that due to
Eq.~(\ref{finitepy}) the values of $\psi^\dagger \psi$ and
$\psi^\dagger \sigma_x\partial_x\psi$ do not depend on the sign
of~$p_y$.

To find the the equilibrium current where all the states with
different sign of $p_y$ are occupied equally we only need to know
the sum $\psi^\dagger_+\sigma_y\psi_+
+\psi^\dagger_-\sigma_y\psi_-$. Using Eq.~(\ref{spincurrent}) to
eliminate the terms $\psi_+^\dagger\sigma_z\psi_+$ and
$\psi_-^\dagger\sigma_z\psi_-$ from Eq.~(\ref{rotatedspin}) we
find
 \bq\label{currentfinal}
\psi^\dagger_{+}\sigma_y\psi_{+} +\psi^\dagger_{-}\sigma_y\psi_{-}
= \fr{-\Delta\hbar v_{\rm F}\, \partial_x(\psi^\dagger_{+}\psi_{+}
+\psi^\dagger_{-}\psi_{-})}{2(\Delta^2 +v_{\rm F}^2p_y^2)}.
 \ee
The {\it r.h.s.} here is effectively a derivative of the electron
density. Substituting Eq.~(\ref{currentfinal}) into
Eq.~(\ref{current_def}) and integrating over $x$ across the
potential step leads to the total AHE current
 \bq\label{current1}
J_y=J_R - J_L \ , \ J_{R(L)} =\fr{-e\Delta\hbar v_{\rm
F}^2}{2}\sum_{i_{R(L)}}
\fr{\psi^\dagger_{i_{R(L)}}\psi_{i_{R(L)}}}{\Delta^2 +v_{\rm
F}^2p_y^2} \ .
 \ee
Since the current density Eq.~(\ref{current_def}) found with the
help of Eq.~(\ref{currentfinal}) is an $x$-derivative, the total
integrated current may be presented as a difference of two
contributions $J_R$ and $J_L$ depending only on the end-points of
integration $x_R$ and $x_L$. In order to find the universal total
current we need to choose these points far to the right and to the
left of the step, where the potential is constant.

If the potential $U(x)$ at the points $x_R$, $x_L$ is flat, one
may choose $\psi_{i_{R(L)}}$ in Eq.~(\ref{current1}) to be single
plane wave solutions of the Dirac equation with either positive or
negative $p_x$ and calculate the sum over $i$ explicitly. Strictly
speaking, any eigenfunction of the Hamiltonian
Eq.~(\ref{HamDirac}) at least on one side of the potential step
$U(x)$ contains both left- ($p_x<0$) and right-moving ($p_x>0$)
waves. Oscillating interference terms between these left- and
right-movers which survive the summation in Eq.~(\ref{current1})
are the only terms carrying the information about the specific
shape of the potential step. These oscillations effectively
average out for $x_R$, $x_L$ taken far away from the step, leading
to an exact AHE current, independent of the shape of the potential
({\it cf.} Fig.~\ref{fig:2} below).

In the case of a slowly varying potential $U(x)$ considered in the
next section, Eq.~(\ref{current1}) allows us to find the part of
the current flowing in a strip $x_L <x <x_R$ even for points
$x_{R(L)}$ inside the step region.

We introduce the valence and conduction band contributions
$J_{R(L)}= J_{R(L)}^c + J_{R(L)}^v$ in Eq.~(\ref{current1}). Then
 \bq\label{current2}
J_{R(L)}^c= \fr{e}{2h}(U_{R(L)}+\Delta -\eps_{\rm F} ) \ ,
 \ee
for $\eps_{\rm F}>\Delta +U_{R(L)}$ and $J_{R(L)}^c=0$ otherwise.
Integration over the momentum direction in Eq.~(\ref{current1}) is
done as $\int_0^{2\pi} {d\phi}/{(a^2 +\cos^2 \phi)} =
{2\pi}/{a\sqrt{a^2 +1}}$. The current Eq.~(\ref{current2}) is
proportional to the difference between the Fermi energy and the
conduction band bottom. To find the contribution to the current
from the valence band electrons, we perform the summation in
Eq.~(\ref{current1}) over all occupied states in that band with
energies higher than some large negative energy $\eps_{\rm min}$.
This gives
 \begin{align}\label{current3}
J_{R(L)}^v= \left\{\begin{array}{cc} \fr{e}{2h} (\eps_{\rm
min}+\Delta-U_{R(L)}) \ , \ \eps_{\rm F}>U_{R(L)}-\Delta \\
\fr{e}{2h}( \eps_{\rm min}-\eps_{\rm F}) \ , \ \eps_{\rm
F}<U_{R(L)}-\Delta
\end{array}\right. .
 \end{align}
The dependance on $\eps_{\rm min}$ disappears in the current $J_y$
defined in Eq.~(\ref{current1}). For electrons with energies
smaller than $\eps_{\rm min}$, the density $\psi^\dagger \psi$ is
constant and the current density, being a derivative of the
particle density, Eq.~(\ref{currentfinal}) vanishes. With the
help of Eqs.~(\ref{current2}, \ref{current3}) we find the total
current as a piecewise linear function of $\eps_{\rm F}$ (see
Fig.~\ref{fig:1})
 \bq\label{current4}
J_y=\left\{\begin{array}{cc} 0 \ , \ \eps_{\rm F} >U_R +\Delta\\
\fr{e}{2h}(\eps_{\rm F}-U_R-\Delta ) \ , \ U_R> \eps_{\rm F}-\Delta> U_L\\
\fr{e}{2h}(U_L-U_R) \ ,  \ U_L+\Delta > \eps_{\rm F} > U_R-\Delta \\
\fr{e}{2h}(U_L-\Delta-\eps_{\rm F}) \ , \ U_R> \eps_{\rm
F}+\Delta> U_L\\
0 \ , \ U_L -\Delta > \eps_{\rm F}
\end{array}\right. ,
 \ee
valid for any shape of the potential step $U(x)$ (we don't even
require monotonous $U(x)$). Generalization of Eq.~(\ref{current4})
for $U_R-U_L>2\Delta$ is presented in the Appendix~A. It is not
surprising that in the central region in Eq.~(\ref{current4}),
where $\eps_{\rm F}$ lies in the gap on both sides of the
potential step and the system is formally an insulator, there
exists a finite constant current. In this paper we consider only
the dissipationless equilibrium AHE current, which for $U_L+\Delta
> \eps_{\rm F} > U_R-\Delta$ is carried by the electrons from the
fully occupied valence band and does not depend on $\eps_{\rm F}$.

It is interesting to compare the AHE current described by
Eq.~(\ref{current4}) to the current which would appear in a
similar setup with the potential $U(x)$ with an additional
quantizing (normal to the $xy$-plane) magnetic field. Due to the
drift of electrons` Larmor orbits transverse to the electric field
each fully occupied Landau level will produce a current in
$y$-direction~\cite{HalperinPRB82} which is twice larger than the
AHE current predicted in Eq.~(\ref{current4}) in the central
region $U_L+\Delta > \eps_{\rm F} > U_R-\Delta$ (the Fermi energy
staying inside the gap at every coordinate $x$). In other words,
we may say that electrons from the fully occupied valence band
produce an AHE current that is exactly one half of the quantum
Hall current due to a single occupied Landau level. For the Fermi
energy below the top of the valence band both to the left and to
the right of the potential step, $U_L -\Delta > \eps_{\rm F}$, the
current $J_y$ in Eq.~(\ref{current4}) disappears which means that
only the electrons near the top of this band contribute to the
AHE. On the other hand, for the Fermi energy being very high in
the conduction band, $\eps_{\rm F}
>U_R +\Delta$, the AHE disappears again. This means that electrons
from the bottom of the conduction band generate an AHE current
which is exactly half of the single Landau level current and has
the opposite sign compared to the valence band current.

The abrupt disappearance of the total AHE current $J_y\equiv 0$,
Eq.~(\ref{current4}), for energy regions $\eps_{\rm F}
>U_R +\Delta$ and $\eps_{\rm F} <U_L -\Delta$ is rather surprising.
A source of the AHE current is attributed to be the Berry
anomalous velocity Eq.~(\ref{Berry}) with $\Omega_z(\vec p)=
\fr{1}{2}\hbar^2\Delta v_{\rm F}^2/(\Delta^2 +p^2 v_{\rm
F}^2)^{3/2}$~\cite{NiuPRL07}. In the case of a monotonous $U(x)$,
the last term in Eq.~(\ref{Berry}) has always the same sign,
meaning there is no current cancellation upon summation over
electronic states. A vanishing $J_y$ in Eq.~(\ref{current4}) may
become possible only because of contributions to the exact current
which are not captured by Eq.~(\ref{Berry}). Such contributions to
the current density, which may be treated semiclassically in the
same manner as Eq.~(\ref{Berry}) are known~\cite{CooperPRB97,
XiaoPRL06}. They are named magnetization currents and appear
because of an inhomogeneity of electron wave-packet rotation. Even
more, we will see in the next section, that the largest
contribution to the microscopic current density comes from the
reflection regions, where the semiclassical approach is not
applicable.

To understand better the spatial distribution of the current
Eq.~(\ref{current1}) together with the role of interference
oscillations between the incoming and outgoing waves, neglected in
Eqs.~(\ref{current2}-\ref{current4}), we describe below a
numerically exact current density in a constant electric field.

\section{III. Current density in a smooth potential}

We choose the potential in Eq.~(\ref{HamDirac}) to be $U(x)=
Fx-\Delta$. In this paper we consider Dirac Hamiltonians with a
gap large enough to neglect tunneling between the bands. The large
insulating gap means that in order to investigate the
AHE current along the border of an electron Fermi liquid it is
enough to take only the conduction band electrons with energies
slightly above the gap into account, {\it i.e.} the
non-relativistic limit $|p_y v_{\rm F}|, |Fx|\ll\Delta$ with the
spinor wave-function (See Appendix B.)
 \bq\label{Airy}
\psi^T=e^{ip_y y}(\phi(x), 0) \ , \ \phi (x) = {\rm
Ai}((x-x_c)/x_0) \ .
 \ee
Here, $\rm Ai(x)$ is the Airy function, $x_0 =(\hbar^2 v_{\rm
F}^2/(\Delta F))^{1/3}$ and $x_c=\eps/F -p_y^2 v_{\rm
F}^2/(2\Delta F)$ and the energy of an electron in the conduction
band is $|\eps|\ll \Delta$.

\begin{figure}
\includegraphics[width=7.1cm]{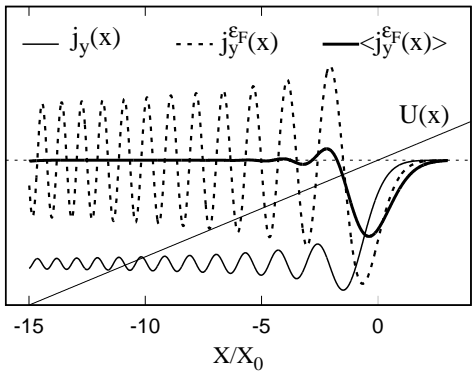}
\caption{\label{fig:2}
 Anomalous Hall current density (conduction
band, $e=-|e|$) for $U(x)=Fx-\Delta$ and $\eps_{\rm F}=0$.
Distances are measured in units of $x_0 =(\hbar^2 v_{\rm
F}^2/(\Delta F))^{1/3}$. The vertical axis has arbitrary units.
The thin solid line is the current density $j_y(x)$. The step-like
current density is formed at distances $\sim x_0$ from the edge.
The dashed line shows the energy distribution of the current
density $j_y^{\eps_{\rm F}}(x)$ and the thick solid line is
$\langle j_y^{\eps_{\rm F}}(x)\rangle$ smoothed with the function
$\exp (-(x-x')^2/x_0^2)$. The clear $\delta$-function character of
$\langle j_y^{\eps_{\rm F}}(x)\rangle$ shows that $j_y^{\eps_{\rm
F}}(x)$ will also behave effectively like a $\delta$-function when
folded with an arbitrary function varying slowly on the scale
$\sim x_0$.
 }
\vspace{-.3cm}
\end{figure}

As we already mentioned, we will from now on consider only the
case of potentials with a relatively small slope, $Fx_0\ll\Delta$.
For steeper potentials, with $F x_0\sim\Delta$, Eq.~(\ref{Airy})
will be no longer valid and one needs to consider the tunneling of
relativistic electrons between the conduction and valence bands.
However, the electrons from the conduction band and from the
valence band generate AHE currents of opposite signs. So the total
AHE is likely to be diminished in the case of tunneling. This may
be seen also from Appendix A, where we consider the extension of
Eq.~(\ref{current4}) (which is valid for arbitrary strong electric
fields $dU/dx$, but where the tunneling is forbidden
energetically) for the case of a large potential step, $U_R-U_L>
2\Delta$, where tunneling is possible.

The current due to a single state Eq.~(\ref{Airy}) is considered
in Appendix B. In Fig.~\ref{fig:2}, we show the total density of
the conduction band current, $j_y(x)$, for $\eps_{\rm F}=0$
obtained after summation over the energy and $p_y$-momentum. The
details of the calculation are given also in Appendix B. Besides
the interference oscillations, the figure agrees well with the
step function form
 \bq\label{currentdensity}
j_y(x)=\fr{e}{2h}F \, \theta(-x) \ .
 \ee
We stress that this result for small $F$ may be derived directly
from Eq.~(\ref{current1}). The current flowing in $y$-direction to
the left of point $x$ is given by Eqs.~(\ref{current1},
\ref{current2}) with $U_R$ replaced by $U(x)$. Differentiating
with respect to $x$ gives Eq.~(\ref{currentdensity}). This
derivation shows that Eq.~(\ref{currentdensity}) is also valid at
large distances, $|Fx|\gtrsim \Delta$.

The authors of Ref.~\cite{LenskyPRL15} considered the AHE for
Dirac electrons in an electric field, but taking into account only
the transport current caused by the Berry velocity
Eq.~(\ref{Berry}). For $|Fx|\ll\Delta$ their result,
$j_y(x)=(e/2h)(x F^2/\Delta)\, \theta(-x)$, is parametrically
smaller than Eq.~(\ref{currentdensity}). It is, however, unclear,
how one can exclusively observe the equilibrium transport current.
Further discussions of the comparison with Ref.~\cite{LenskyPRL15}
are given in Appendix B.

Features of the semiclassical intrinsic AHE for massive Dirac
Hamiltonians are best revealed by the energy distribution of the
current density $\vec j^\eps(\vec r)$, also considered in
Ref.~\cite{CooperPRB97} and defined as $\vec j(\vec
r)=\int^{\eps_{\rm F}}\vec j^\eps(\vec r)d\eps$. Numerical plots
in Fig.~\ref{fig:2} show that $\vec j^\eps (\vec r)$ in a uniform
electric field is mathematically equivalent to the
$\delta$-function in the $\hbar\rightarrow 0$ limit. This means
that we may expect that $\vec j^\eps(\vec r)$ will be concentrated
along the lines $U(\vec r)=\eps-\Delta$ in the case of an
arbitrary slowly varying two-dimensional potential also. Thus, we
can write 
 \bq\label{distrib_current}
\vec{j}^\eps(\vec r) =\fr{e}{2h}\,\, \delta(\eps -U(\vec
r)-\Delta) \,\, \vec n_z\times\vec{\nabla}U(\vec r)
 \ .
 \ee
This formula is the central result of our paper. The exact shape
of the $\delta$-function may be deduced from
Eqs.~(\ref{current_appendix}, \ref{current_distr_S}) of Appendix B
 \bq\label{distrib_delta}
\delta(\eps -U(\vec r)-\Delta)|\vec{\nabla}U|\rightarrow
\fr{d}{dr_\perp}\int_0^\infty {\rm Ai}^2(\xi^2+\fr{r_\perp}{r_0})
d\xi \ ,
 \ee
where $r_\perp =(\eps -U(\vec r)-\Delta)/|\vec{\nabla}U|$ is the
distance between $\vec r$ and the equipotential line $U(\vec
r)=\eps -\Delta$ and ${\rm Ai}$ is again the Airy function. The
width of the $\delta$-function, nonperturbative both in $\hbar$
and in the electric field, is $r_0=(\hbar^2 v_{\rm F}^2/(\Delta
|\vec \nabla U(\vec r)|))^{1/3}$.

The shape of the $\delta$-function Eq.~(\ref{distrib_delta}) is
illustrated in Fig.~\ref{fig:2}. At first glance, the curve shown
in the figure does not seem a good approximation for the
$\delta$-function due to its large oscillations. Nevertheless, the
period of these oscillations decreases fast with increasing
$r_\perp$ \ (or $x$ in the figure) and in the limit
$r_0\rightarrow 0$ ($x_0\rightarrow 0$) it satisfies the property
of the functional $\int \delta (x)f(x)dx =f(0)$ for any smooth
$f(x)$.

Equation~(\ref{distrib_current}) is valid if $|\partial^2
U/\partial{r_i}^2|\, r_0\ll |\vec \nabla U|$. Absence of
inter-band tunneling requires $r_0|\vec\nabla U|\ll\Delta$. A
simple direct proof of Eq.~(\ref{distrib_current}) in the
nonrelativistic limit is given in Appendix C.

The $\delta$-function Eq.~(\ref{distrib_current}) is peaked at the
border of the area accessible classically at the energy $\eps$.
This defines the line of stopping points, where both components of
the momentum of an electron reaching such a point vanish. Inside
this area $\vec{j}^\eps(\vec r)=0$. (see also Fig.~\ref{fig:5_s}
and the discussion in Appendix B.)

For the valence band Eq.~(\ref{distrib_current}) changes sign. The
total current density $j^c_y(x)+j^v_y(x)$ vanishes everywhere
where the Fermi energy stays locally inside the conduction or the
valence band. For example, in the case of a sufficiently strong
and smooth disorder potential the Fermi energy may cross several
times both the bottom of the conduction and the top of the valence
bands. The sample in this case, even being insulating on average,
will consist of large electron and hole puddles separated by the
big insulating regions. Our theory in this case predicts the
vanishing AHE current inside the puddles and existence of the AHE
in the insulating parts between the puddles.

\section{IV. Ray dynamics and magnetization currents}

The eigenfunction of the Hamiltonian Eq.~(\ref{HamDirac}) in the
semiclassical limit may be written in the form (we use $\hbar
=v_{\rm F}=1$ in this section)
 \bq\label{psiab}
\psi(x) = \sqrt{1/v_x}\, e^{i\int^x p(x')dx'} [1 +\beta(x)\sigma_y
] \phi\ .
 \ee
Here, $\phi^T(x)=\fr{1}{\sqrt{2}}\left(\sqrt{1+\xi },
\sqrt{1-\xi}\right)$, $\xi = \Delta/\sqrt{\Delta^2 +p^2}$ and $
p(x) =\sqrt{(\eps-U(x))^2-\Delta^2} $. For illustrative purposes
we only consider the case of an incident ray parallel to the
potential gradient. The classical longitudinal velocity of
electrons in the ray is $v_x =p/\sqrt{\Delta^2+p^2}=p/(\eps -U)$.
(Arbitrary incident angles may be considered with the help of
Eq.~(\ref{finitepy}))

The coefficient $\beta$ in Eq.~(\ref{psiab}), responsible for the
expectation value of the anomalous velocity, may be found by
acting with $\psi(x)$ on the Hamiltonian Eq.~(\ref{HamDirac}), as
is shown in Appendix D. Alternatively, the value of $\beta$ may be
extracted in the linear approximation in $U'=dU/dx$ from the exact
Eq.~(\ref{currentfinal}) in which we substitute the wave function
Eq.~(\ref{psiab}). Doing so, we find
 \bq\label{v_Magnet}
\langle v_y\rangle=
\fr{\psi^\dagger\sigma_y\psi}{\psi^\dagger\psi}
=2\beta=\fr{-U'\Delta}{2 p^2\sqrt{\Delta^2+p(x)^2}} \ .
 \ee
This velocity is bigger, and even much bigger if $p(x)\ll \Delta$,
than the AHE velocity deduced from Eq.~(\ref{Berry}). However,
Eq.~(\ref{v_Magnet}) does not provide the true information about
the transverse transport, since the solution $\psi(x)$ extends
indefinitely along the $y$-axis. Even more, as we show below, the
term $\sim\beta$ in the wave function Eq.~(\ref{psiab}) simply
does not contribute to the electrons' trajectory.

To find the actual bending of the trajectory, we need to consider a
ray of electrons
 \bq\label{ray1}
\Psi(x,y)=\int f(p_y)\psi_{p_y}(x,y) dp_y \ ,
 \ee
where $\psi_{p_y}(x,y)$ are the solutions of the Dirac equation
Eq.~(\ref{HamDirac}) with finite momentum $p_y$ along the
potential step, having all the same energy $\eps$ and the narrow
function $f(p_y)$ is peaked at $p_y=0$.

\begin{figure}
\includegraphics[width=8.25cm]{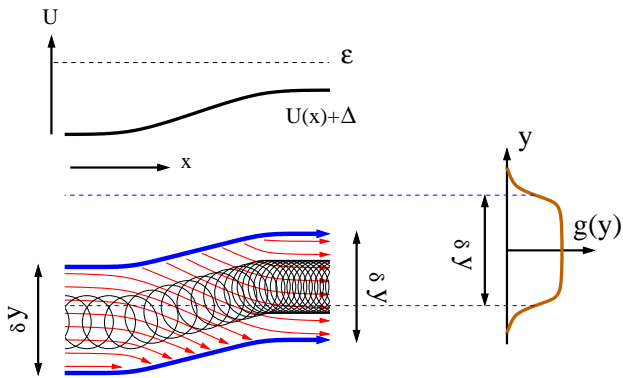}
\caption{\label{fig:3} The electron ray with width $\delta y$
traversing the region of a smooth potential step $U(x)$ with
energy $\eps
>\Delta+U$. Blue lines are the borders of the ray with
approximately constant electron density inside (envelope function
$g(y)$). Red lines show the velocity field $\langle \vec v \rangle
=\langle \vec\sigma\rangle$. There is a side jump of the ray
during crossing the electric field area, $dU/dx\neq 0$, consistent
with the anomalous velocity $\vec v_B=\vec\Omega\times \vec\nabla
U/\hbar$, Eq.~(\ref{Berry}). According to Eqs.~(\ref{PsiShifted},
\ref{y(x)}) the ray is always shifted downwards. The magnitude of
the shift is bigger for larger velocity $v_x$, or, equivalently,
for larger $\eps-U(x)$. Circles inside the ray show how the
nonuniform electron's magnetization results in a current. The
density of the circles (magnetization), rotating
counter-clockwise, increases to the right on the figure.
 }
\end{figure}

For small $p_y$, Eq.~(\ref{finitepy}) gives $\psi_{p_y}(x,y) =
e^{ip_y y}(1+i\sigma_x {p_y}/{(2\Delta)})\psi(x)$ and
Eq.~(\ref{ray1}) becomes
 \bq\label{ray2}
\Psi =\left(1+\fr{\sigma_x}{2\Delta}\fr{\partial}{\partial
y}\right)g(y)\psi(x) \ ,
 \ee
where $g(y)=\int dk e^{iky}f(k)$ is a smooth envelope function in
$y$ direction of a ray propagating mostly along the $x$-axis. It
is convenient to choose $g(y)$ almost flat within some region
$\delta y\gg 1/p(x)$ which smoothly goes to zero outside the
region, see Fig.~\ref{fig:3}.

To find explicitly the transverse displacement of the ray we
rewrite Eq.~(\ref{ray2}) as
 \bq\label{PsiShifted}
\Psi =g\left(y+\fr{1}{2\Delta}\right)\psi_+(x)
+g\left(y-\fr{1}{2\Delta}\right)\psi_-(x)\ ,
 \ee
where we use the decomposition of $\psi(x)$ into the $\sigma_x$
eigenvectors, $\psi =\psi_+ +\psi_-$ and $\sigma_x\psi =\psi_+
-\psi_-$. The trajectory $y(x)$ is now found as (see
Fig.~\ref{fig:3})
 \bq\label{y(x)}
y(x)
=\fr{1}{2\Delta}\fr{-\psi^\dagger(x)\sigma_x\psi(x)}{\psi^\dagger(x)
\psi(x)}=\fr{-p(x)}{2\Delta\sqrt{\Delta^2+p(x)^2}} \ .
 \ee
The shift of the trajectory $y(x)$ exists already for the wave
function Eq.~(\ref{psiab}) in zeroth order in $U'$ when the term
$\sim\beta$ is omitted. The transverse transport velocity is now
 \bq\label{v_y}
v_{y{\rm tr}}=v_x\fr{dy}{dx}=
\fr{U'\Delta}{2(\Delta^2+p(x)^2)^{3/2}}\ ,
 \ee
in accordance with Eq.~(\ref{Berry}).

Both the distribution of the local velocity $\langle \vec
v\rangle$ (with the $y$-component $\langle \vec v_y\rangle$ given
by Eq.~(\ref{v_Magnet})) and the shift of the ray $y(x)$
Eq.~(\ref{y(x)}) are shown in Fig.~\ref{fig:3}. It is hard to show
in the figure what happens in the narrow region at the borders of
the ray, where the red lines of the velocity field cross the ray
border shown in blue. An accurate description of the charge
balance at such borders is presented in Appendix E, where it gives
yet another way to find the anomalous velocity
Eq.~(\ref{v_Magnet}) and the coefficient $\beta$, see
Eq.~(\ref{jyvy supp}).

The total current is the sum of the transport part ($\vec j_{tr}$)
and a magnetization contribution ($\vec\nabla\times\vec M(\vec r)$
with $\vec M(\vec r)$ being the magnetic moment
density)~\cite{CooperPRB97},
 \bq\label{Magnetization_Halperin}
\vec j=\vec j_{tr} +\vec\nabla\times\vec M(\vec r) \ .
 \ee
In Appendix E we use $M(\vec r)$ available in the
literature~\cite{NiuPRL07} to show that our results
Eqs.~(\ref{v_Magnet}, \ref{v_y}) indeed agree with
Eq.~(\ref{Magnetization_Halperin}). For an electron in the valence
band $\vec M(\vec r)$ is larger in the regions with higher
potential $U(x)$. Consequently, in Fig.~\ref{fig:3}, we illustrate
the magnetization current $\vec\nabla\times\vec M(\vec r)$,
Eq.~(\ref{Magnetization_Halperin}), by drawing circles with a
coordinate dependent density.

The transport (Eq.~(\ref{v_y})) and the total
(Eq.~(\ref{v_Magnet})) anomalous Hall currents in Fig.~\ref{fig:3}
flow in opposite directions. However, the figure shows the
semiclassical current Eq.~(\ref{v_Magnet}) due to a single
injected electron ray. In the case of the equilibrium AHE there
will be many such rays, including the ones reflected by the
potential. The total equilibrium transverse current~--~flowing in
the direction suggested by Eq.~(\ref{Berry})~--~originates from
the narrow reflection regions, Eq.~(\ref{distrib_current}), not
captured by Eqs.~(\ref{psiab}, \ref{v_Magnet}). This shows again
the importance of our results that go beyond the semiclassical
treatment.

Considering an electron ray instead of a wave
packet~\cite{MacDonaldRMP10,NiuRMP10} allows us at once to find
the entire trajectory Eq.~(\ref{y(x)}). Interestingly, the
anomalous shift of the trajectory Eqs.~(\ref{PsiShifted},
\ref{y(x)}) turned out to have an exact upper bound,
$|y|<1/(2\Delta)$.

\section{V. Conclusions}

In this paper, we calculated the total local AHE current for
electrons described by the massive Dirac Hamiltonian. The exact
results turned out to be strongly universal in the case where the
potential $U(x)$ depends only on one coordinate. What is even more
surprising, the current which we found for an arbitrarily smooth
potential $U(\vec r)$ turns out to be much stronger (and its
energy/coordinate dependance much sharper) than the usually
considered Berry curvature-induced currents. For example, the
equilibrium anomalous Hall currents exist if the Fermi energy lies
inside the insulating gap, but disappear abruptly if $\eps_{\rm
F}$ is shifted into the conduction or valence band. The width of
the transition is governed by the weakness of the electric field
or the size of the sample. It will be interesting to see this
sharp Fermi energy dependance in measurements of the quantum Kerr
and Faraday effects for the surface states in three-dimensional
topological insulators~\cite{FaradayExp}.

Another promising direction for further research lies in the
understanding of the relations between the non-dissipative
equilibrium bulk AHE currents considered here and the protected
edge currents found in the topological
insulators~\cite{inpreparation}

{\it Acknowledgements.--} Discussions with  Sunghun Park,
P.~W.~Brouwer, C.~W.~J.~Beenakker, I.~V.~Gornyi and  C.~De~Beule
are greatly acknowledged. This work was supported by the DFG grant
RE~2978/8-1.



\bigskip

\section{Appendix A: Anomalous Hall current for the case $U_R-U_L
>2\Delta$}

Here, we consider the generalization of the formula
Eq.~(\ref{current4}) to the case of a large potential step $U(x)$
with a magnitude that exceeds the value of the gap in the electron
spectrum, $U_R-U_L
>2\Delta$.

As in the main text, we assume that far away from the step the
potential $U(x)$ became flat with the asymptotic values $U_R,
U_L$.
That means we may still
perform explicitly a summation over all occupied states in
Eq.~(\ref{current1}) to obtain the same Eqs.~(\ref{current2},
\ref{current3}). Integration over the momentum direction in
Eq.~(\ref{current1}) is done as $\int_0^{2\pi} {d\phi}/{(a^2
+\cos^2 \phi)} = {2\pi}/{a\sqrt{a^2 +1}}$. Rewritten in a more
detailed form, Eqs.~(\ref{current2}, \ref{current3}) become (note
that the electron's charge is negative, $e=-|e|$)
 \begin{align}\label{current2S}
&J_{R}^c= \left\{\begin{array}{cc} \fr{e}{2h}(U_{R} +\Delta
-\eps_{\rm
F}) \ , \ \eps_{\rm F}>U_{R}+\Delta \\
 0 \ , \ \eps_{\rm F}< U_{R}+\Delta
\end{array}\right. , \nonumber\\
&J_{L}^c= \left\{\begin{array}{cc} \fr{e}{2h}(U_{L} +\Delta
-\eps_{\rm
F}) \ , \ \eps_{\rm F}>U_{L}+\Delta \\
 0 \ , \ \eps_{\rm F}< U_{L}+\Delta
\end{array}\right. ,
 \end{align}
and
 \begin{align}\label{current3S} &J_{R}^v=
\left\{\begin{array}{cc}
 \fr{e}{2h}(\eps_{\rm min}+\Delta-U_R) \ , \ \eps_{\rm F}>U_{R}-\Delta \\
\fr{e}{2h}(\eps_{\rm min} -\eps_{\rm F}) \ , \ \eps_{\rm
F}<U_{R}-\Delta
\end{array}\right. , \nonumber\\
&J_{L}^v= \left\{\begin{array}{cc}
\fr{e}{2h}(\eps_{\rm min}+\Delta-U_L) \ , \ \eps_{\rm F}>U_{L}-\Delta \\
\fr{e}{2h}(\eps_{\rm min} -\eps_{\rm F} ) \ , \ \eps_{\rm
F}<U_{L}-\Delta
\end{array}\right. .
 \end{align}
As we explained in the main text, $\eps_{\rm min}$ is the lower
bound of the integral(sum) over energies in Eq.~(\ref{current1})
which cancels out from the current $J_y$. Electrons with energies
below $\eps_{\rm min}$ do not contribute to the current since for
them the derivative of the density $\partial_x\psi^\dagger\psi$
vanishes in Eq.~(\ref{currentfinal}). We assume that there is no
contribution to AHE from the bottom (ultraviolet cutoff) of the
valence band. The formal proof of the latter will be given
elsewhere~\cite{inpreparation}.

\begin{figure}
\includegraphics[width=5.8cm]{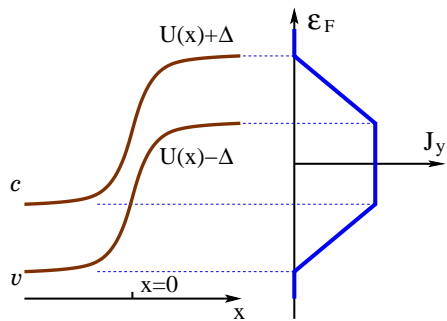}
\caption{\label{fig:4_s} Similar to Fig.~\ref{fig:1}, but for
$U_R-U_L
>2\Delta$: total anomalous Hall current $J_y$ flowing along the
potential step $U(x)$ as a function of the Fermi energy $\eps_{\rm
F}$ for the two-dimensional massive Dirac Hamiltonian with a
spectral gap $2\Delta$.
 }
\vspace{-.3cm}
\end{figure}

The band structure in the presence of a large potential step is
shown in Fig.~\ref{fig:4_s} and needs to be compared to
Fig.~\ref{fig:1}. The five regimes of different current behaviors
$J_y(\eps_{\rm F})$ in Fig.~\ref{fig:1} represent (from lower to
higher energies $\eps_{\rm F}$) metallic, half-metallic,
insulating, half-metallic and metallic phases. Here, metallic
refers to $\eps_{\rm F}$ lying inside the conduction or valence
band, insulating means that $\eps_{\rm F}$ always lies inside the
gap and half-metallic means that $\eps_{\rm F}$ is inside the gap
only on one side (half plane) of the potential step. The sequence
of regimes in Fig.~\ref{fig:4_s} is metallic, half-metallic,
metal-metal, half-metallic and metallic where the metal-metal
regime corresponds to two half-plane metals separated by
a tunnel barrier. The total anomalous Hall current for the setup
in Fig.~\ref{fig:4_s} found from Eqs.~(\ref{current1},
\ref{current2S}, \ref{current3S}) is
 \bq\label{current4S}
J_y=\left\{\begin{array}{cc} 0 \ , \ \eps_{\rm F} >U_R +\Delta\\
\fr{e}{2h}(\eps_{\rm F}-U_R -\Delta) \ , \ U_R+\Delta > \eps_{\rm
F}> U_R -\Delta\\
-\fr{e}{h}\Delta \ ,  \ U_R-\Delta > \eps_{\rm F} > U_L+\Delta \\
\fr{e}{2h}(U_L -\Delta -\eps_{\rm F}) \ , \ U_L+\Delta> \eps_{\rm
F}> U_L-\Delta\\
0 \ , \ U_L -\Delta > \eps_{\rm F}
\end{array}\right. ,
 \ee
valid again for any shape of the potential step $U(x)$. The main
difference to Eq.~(\ref{current4}) is the smaller current in the
central metal-metal regime, i.e. $J_y=-\fr{e}{h}\Delta$ instead of
$J_y=\fr{e}{h}(U_L -U_R)$.

\section{Appendix B:  current density in an uniform
electric field}

Like in the main text, we consider here a linear potential
$U(x)=Fx-\Delta$ and assume the nonrelativistic limit of the
conduction band electrons, i.e. $|Fx|\ll \Delta$, $|p_y v_{\rm
F}|\ll\Delta$. The Dirac equation Eq.~(\ref{HamDirac}) for the
spinor wave function $\psi^T=(\phi,\chi)$ reduces in this case to
an non-relativistic Schr\"{o}dinger equation for the upper
component, $\phi$, with the lower component being negligibly
small, $|\chi|\ll|\phi|$,
 \bq
\left(\fr{p^2}{2m} +Fx\right)\phi =\eps\phi \ , \ \chi
=\fr{p_x+ip_y}{2mv_{\rm F}}\phi \ .
 \ee
Here, the mass $m$ is related to the gap parameter via $\Delta =m
v_{\rm F}^2$. The Schr\"{o}dinger equation for $\phi$ is solved by
the Airy function as used in Eq.~(\ref{Airy}) of the main text. In
this limit, the anomalous Hall current density
Eqs.~(\ref{current1}, \ref{current2}) becomes proportional to the
derivative of the electron charge density
 \bq\label{currentNonRS}
j_y(x)= -e\fr{\hbar}{2m} \partial_x \rho(x) \ .
 \ee
Also without the loss of generality we may put Fermi energy equal
to zero, $\eps_{\rm F}=0$.

\subsection{Current for a single electron}

\begin{figure}
\includegraphics[width=7.8cm]{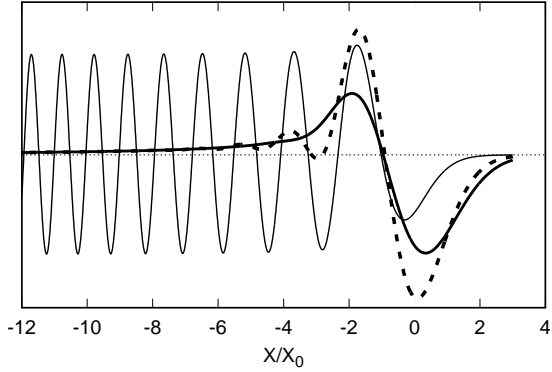}
\caption{\label{fig:5_s} Anomalous Hall current density $j_y(x)$
for a single electron (conduction band) with energy $\eps=0$ and
$p_y=0$ for the linear potential $U(x)=Fx-\Delta$. Distances are
measured in units of $x_0 =(\hbar^2 v_{\rm F}^2/(\Delta
F))^{1/3}$. The thin solid line shows the current density
Eq.~(\ref{currentNonRS}) with strong oscillations having constant
amplitude at negative $x$. To reveal the smooth semiclassical
contribution to the current density we show the plot of the same
current smoothed with weights $\exp (-((x-x')/x_0)^2)$ (dashed
line) and $\exp (-((x-x')/x_0)^2/2)$ (thick solid line). We do not
specify the units at the vertical axis, but the relative amplitude
for the smoothed curves is multiplied by a factor of $4$.
 }
\vspace{-.3cm}
\end{figure}

We start by considering the current density
Eq.~(\ref{currentNonRS}) contributed by a single electronic state
described by the wave function Eq.~(\ref{Airy}). The result of
such a calculation, shown in Fig.~\ref{fig:5_s} for $p_y=\eps=0$,
demonstrates strong interference oscillations between incoming and
reflected electron waves. As follows from the asymptotic form of
the Airy function for large negative $x$, the oscillation
amplitude stays constant but its period decreases like $\sim
1/|x|^{3/2}$.

In order to visualize the semiclassical part of the current
density we smooth out the interference oscillations by averaging
the density at each point $x$ with the weight functions
$e^{-((x-x')/x_0)^2}$ and $e^{-((x-x')/x_0)^2/2}$ in
Fig.~\ref{fig:5_s}. Assuming an oscillation amplitude of unity,
the semiclassical current density at large negative $x$ is small
like $1/(4|x|^{3/2})$. Therefore, we have to enlarge the smoothed
current density in the figure in order to make the semiclassical
current density visible.

The smoothed current density shown in Fig.~\ref{fig:5_s} (thick
solid line) has two distinct features. The first was already
mentioned. It is the positive power law tail of $j_y(x)$ for large
negative values of $x$. (Only at this tail the current density may
be explained by Eq.~(\ref{v_Magnet}) for the special case of a
weak linear potential.) The second is the large negative
($e=-|e|$) bump of the current density at the reflection point
$x=0$. The current density, Eq.~(\ref{currentNonRS}), is
proportional to the derivative of the electron density, which in
the semiclassical limit is $\rho(x)\sim 1/\sqrt{-x}$ at large
negative $x$ and $\rho(x)\approx 0$ at positive $x$. That is why
both the power law tail of $j_y(x)$ and the negative bump are
inevitable and survive the smoothing procedure. The total
single-electron current integrated over $x$ vanishes in the
non-relativistic approximation Eq.~(\ref{Airy}). The negative bump
of the current density at the reflection point is responsible for
the vanishing of the current density energy distribution $\vec
j^\eps (\vec r)$ inside the classically accessible area, see
Eq.~(\ref{distrib_current}). (Note that $\vec j^\eps (\vec r)$ is
a sum of currents due to many electrons with the same energy
$\eps$, while Fig.~\ref{fig:5_s} shows only the single electron
current.)

\subsection{Summing up the current contributions due to many electrons}

The semiclassical electron density
in the conduction band is given by the integral
 \bq\label{density_classic}
\rho =\int_{\eps<\eps_{\rm F}} \fr{d^2p}{(2\pi\hbar)^2} =
\fr{2}{(2\pi\hbar)^2} \int \fr{d\eps_x dp_y}{d\eps_x/dp_x} \ ,
 \ee
where $\eps_x =p_x^2/(2m)$ and the coordinate-dependance emerges
through the limits of integration, $\eps=\vec p^2/2m +Fx <
\eps_{\rm F}$. To make a connection between this formula and the
wave function Eq.~(\ref{Airy}) we notice that the smooth part of
the squared Airy function, representing the semiclassical density,
using the negative $x$ asymptotics may be written as (here
$p_y=0$, generalization for finite $p_y$ is obvious, see
Eq.~(\ref{density_appendix}) below)
 \bq
\langle {\rm Ai}^2\left(\fr{x}{x_0}\right)\rangle
=\fr{\sqrt{x_0}}{2\pi\sqrt{-x}}= \fr{\sqrt{2mFx_0}}{2\pi p_x(x)}=
\fr{\sqrt{2Fx_0/m}}{2\pi d\eps_x/dp_x} \ ,
 \ee
where we have used $p_x(x)=\sqrt{-2mFx}$ and $\langle...\rangle$
denotes the average procedure. Consequently, we may replace the
semiclassical density of the conduction band electrons
Eq.~(\ref{density_classic}) by the exact formula
 \begin{align}\label{density_appendix}
&\rho(x)= \fr{1}{\pi \hbar^2\sqrt{2Fx_0/m}} \\
\times & \int_{\eps_x >0} {\rm Ai}^2\left(
\fr{\eps_x+p_y^2/2m+Fx}{Fx_0} \right)d\eps_x dp_y \ . \nonumber
 \end{align}
The absence of electrons with energies above the Fermi energy,
$\eps_{\rm F}=0$, is taken care of automatically due to the
exponential suppression of the Airy function for positive arguments,
so there is no need to introduce an upper limit of integration
over either $\eps_x$ or $|p_y|$.

With the help of Eq.~(\ref{currentNonRS}) we may find the current
density
 \begin{align}\label{current_appendix}
&j_y(x)= \fr{e F}{2\pi m\hbar\sqrt{2Fx_0/m}} \\
\times & \int_{\eps_y>0} {\rm Ai}^2\left( \fr{\eps_y+Fx}{Fx_0}
\right)dp_y \ . \nonumber
 \end{align}
This formula is used for calculating the current density in
Fig.~\ref{fig:2}. The easiest way to find the energy distribution
of the current density $j^\eps_y(x)$ is by differentiating
Eq.~(\ref{current_appendix}) to obtain
 \bq\label{current_distr_S}
j_y^{\eps=0}(x)= \fr{1}{F}\fr{\partial}{\partial x} j_y(x) \ .
 \ee
This result is shown in Fig.~\ref{fig:2}.

Replacing the squared Airy function by its asymptotic behavior we find
the semiclassical density of the anomalous Hall current (valid only for
negative $x$)
 \bq\label{current_step_S}
j_y(x) =\fr{eF}{4\pi^2\hbar} \int_0^{-Fx}
\fr{d\eps_y}{\sqrt{\eps_y(|Fx|-\eps_y)}} \ .
 \ee
Upon energy integration we arrive at Eq.~(\ref{currentdensity}),
$j_y(x)\propto \theta (-x)$.

\begin{figure}
\includegraphics[width=7.7cm]{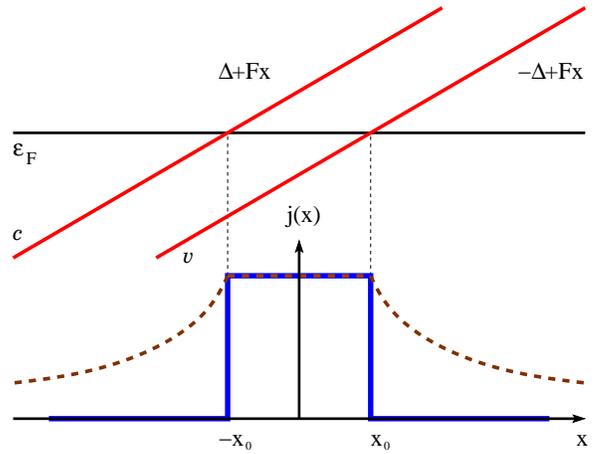}
\caption{ \label{fig:4} {\it Top -} Energy bands structure. {\it
Bottom -} Comparison of the density of the anomalous Hall current
found in this paper (solid-blue) with the current density due to
side jumps only~\cite{LenskyPRL15} (dashed-brown) around the band
crossing by the Fermi energy in a strong electric field.
 }
\end{figure}

\subsection{Comparison to existing results}

Our results for the anomalous Hall current density for the massive
Dirac Hamiltonian (e.g. Eq.~(\ref{currentdensity}) of the main
text) differ substantially from the calculation of dissipationless
bulk currents in a recent paper Ref.~\cite{LenskyPRL15} where only
the transport contribution to the anomalous Hall current caused by
the Berry velocity in Eq.~(\ref{Berry}) was taken into account.
For the potential $U=Fx$ (shifted by $\Delta$ from what we used
before) and the Fermi energy $\eps_{\rm F}=0$ the authors of
Ref.~\cite{LenskyPRL15} have found the total (conduction plus
valence band) density of anomalous current in the form
 \bq\label{Levitov_current}
j_y(x)=\left\{\begin{array}{cc} 
j_0 \ &, \ |x|<x_0\\
j_0 x_0/|x|\ &, \ |x|>x_0
\end{array}\right. \ \ , \ \ j_0=\fr{|e| F}{2h} \ .
 \ee
Here $x_0=\Delta/F$ and $x=\pm x_0$ are the classical turning
points for the conduction and valence band electrons at the Fermi
energy. (Note that the total current $\int j_y(x)dx$ diverges
logarithmically.)

Our approach for a sufficiently small $F$ would give instead of
Eq.~(\ref{Levitov_current}) the current density
 \bq\label{non_Levitov_current}
j_y(x)=\left\{\begin{array}{cc} 
j_0 \ &, \ |x|<x_0\\
0 \ &, \ |x|>x_0
\end{array}\right. \ \ , \ \ j_0=\fr{|e| F}{2h} \ .
 \ee
The current density is constant and proportional to the electric
field only when the Fermi energy lies in the gap between two
bands.

The results Eq.~(\ref{Levitov_current}) and
Eq.~(\ref{non_Levitov_current}) are compared in Fig.~\ref{fig:4}.
As we mentioned already, the reason for the difference between
Eqs.~(\ref{Levitov_current}) and (\ref{non_Levitov_current}) is
that the authors of Ref.~\cite{LenskyPRL15} calculate only the
transport current caused by the Berry velocity contribution
Eq.~(\ref{Berry}) to the motion of the center of the wave packet.
Our derivation, starting from the calculation of the expectation
value of the velocity operator automatically includes both the
motion of its center and the inhomogeneous rotation of the
wave-packet, creating the magnetization current. Although it was
argued~\cite{CooperPRB97, XiaoPRL06} that the magnetization
current is irrelevant for transport phenomena, it is necessary for
finding {\it e.g.} the magnetic moment of the electron gas.


Taking into account only the conduction band contribution to the
current near the Fermi energy crossing with the bottom of the
conduction band in Eq.~(\ref{non_Levitov_current}) gives
 \bq\label{currentdensityS}
j_y(x)=\fr{e}{2h}F \, \theta(-x_0-x) \ ,
 \ee
which is the same as Eq.~(\ref{currentdensity}) with shifted
coordinate due to a different definition of the constant electric
field potential. Analogously, extracting the conduction band
current from Eq.~(\ref{Levitov_current}) at $x\approx -x_0$ gives
(for the Fermi energy above the top of the valence band and only
slightly above the bottom of the conduction band, valence band
electrons give a large constant contribution to the current, while
the linear in $x$ (at $|x-x_0|\ll x_0$) part of $j_y(x)$ comes
from the conduction band electrons)
 \bq\label{currentdensityS_Levitov}
j_y(x)=\fr{-e}{2h}\fr{(x_0+x)F^2}{\Delta} \theta(-x_0-x) \ ,
 \ee
which is parametrically smaller than our result. The linear
increase of the current in Eq.~(\ref{currentdensityS_Levitov})
reflects the fact that the electron density in a two-dimensional
non-relativistic electron gas in a constant electric field
increases linearly with the coordinate. All these electrons have
the same anomalous velocity Eq.~(\ref{Berry}). The current density
of Eq.~(\ref{currentdensityS}) is much more singular than
Eq.~(\ref{currentdensityS_Levitov}) and leads to the
$\delta$-function energy distribution of the current density $\vec
j^\eps (\vec r)$, Eq.~(\ref{distrib_current}).

\subsection{Vanishing of $\vec j^\eps (\vec r)$}

Integration over energy
in Eq.~(\ref{current_step_S}) gives a step-shaped current density
Eq.~(\ref{currentdensity}). Differentiating this result like in
Eq.~(\ref{current_distr_S}) gives a $\delta$-function energy
distribution of the current density $j_y^\eps(x)$. Vanishing of
$j_y^\eps(x)$ almost everywhere except the close vicinity of the
line $\Delta +U(x)- \eps=0$ allows us to suggest a general
$\delta$-function formula for the current in arbitrary smooth
potential $U(\vec r)$, Eq.~(\ref{distrib_current}).

The $\delta$-function Eq.~(\ref{distrib_current}) peaks at the
border of the area accessible classically at the energy $\eps$.
This is the line of stopping points where both components of
momentum of an electron reaching the point vanish. Inside this
area $\vec{j}^\eps(\vec r)=0$, which is somewhat surprising since
the semiclassical current density (given by the $\sim 1/|x|^{3/2}$
tail of the smoothed density in Fig.~\ref{fig:5_s}) for each
electron is always of the same sign. The only current of the
"wrong" sign, ensuring the vanishing of the current density
distribution, is the negative bump in the smoothed current density
in Fig.~\ref{fig:5_s} at the classical turning point. Through
every point $\vec r$ there exist two trajectories with energy
$\eps$ and momentum $\vec p$ normal to the local electric field.
These are trajectories having a turning point in the sense of
Fig.~\ref{fig:5_s} and the "negative bump" in the current density
at~$\vec r$.

\section{Appendix C: Explicit proof of the formula for $\vec
j^\eps(\vec r)$.}

In this Appendix we first give a proof of
Eq.~(\ref{distrib_current}) in the nonrelativistic limit by
calculating the current carried by the conduction band electrons
for a smooth two-dimensional potential and a Fermi energy only
slightly above the insulating gap.

Consider the Dirac Hamiltonian
 \bq\label{HamDiracApC}
H=v_{\rm F}(\sigma_x p_x +\sigma_y p_y) +\Delta\sigma_z +U(\vec r)
-\Delta \ .
 \ee
Calculation of the anomalous Hall current (carried e.g. by the
conduction band electrons) becomes especially easy when $|U(\vec
r)|\ll\Delta$.

Let the eigenfunction of $H$ in Eq.~(\ref{HamDiracApC}) have the
form
 \bq\label{phichi}
\psi=\left(\begin{array}{cc}
\phi\\
\chi
\end{array}\right) \ .
 \ee
Substituting this into Eq.~(\ref{HamDiracApC}) in the limit
$|\eps|\ll \Delta$, $|U|\ll\Delta$, one readily recovers the
nonrelativistic Schr\"odinger equation
 \bq\label{Schroed_supp}
\left(\fr{p^2}{2m} + U(\vec r)\right)\phi =\eps\phi \ , \ \chi =
 \fr{-i\hbar\partial_x +\hbar\partial_y}{2mv_{\rm F}}\phi \ .
 \ee
Here, $\Delta = m v_{\rm F}^2$. We refer to the electron state
described by the Dirac Hamiltonian Eq.~(\ref{HamDiracApC}) with at
least one component of momentum effectively comparable to
$\Delta/v_{\rm F}$ as relativistic and the electron described by
the Eq.~(\ref{Schroed_supp}) as nonrelativistic. The two
components of the current $\vec j =ev_{\rm F}\langle \vec\sigma
\rangle$ are now found as (compare to Eq.~(\ref{currentfinal}))
 \begin{align}\label{psipsi_supp}
\psi^\dagger\sigma_y\psi = -\fr{\hbar v_{\rm
F}}{2\Delta}\partial_x \phi^*\phi &-\fr{i\hbar v_{\rm
F}}{2\Delta}[\phi^*\partial_y\phi-(\partial_y\phi^*)\phi] \ ,
\nonumber\\
\psi^\dagger\sigma_x\psi = \fr{\hbar v_{\rm F}}{2\Delta}\partial_y
\phi^*\phi  &-\fr{i\hbar v_{\rm
F}}{2\Delta}[\phi^*\partial_x\phi-(\partial_x\phi^*)\phi] \ .
 \end{align}
Here, the second terms on the r.h.s. of both equations have the
form of the usual currents in non-relativistic quantum mechanics,
whereas anomalous Hall current is given by the first terms. The
electron density in the semiclassical and non-relativistic
approximation is
 \bq\label{nr_density_supp}
\sum_{\eps_i<\eps_{\rm F}} \psi_i^\dagger
\psi_i=\int_{\fr{p^2}{2m}+U<\eps_{\rm F}} \fr{d^2
p}{(2\pi\hbar)^2} = \theta(\eps_{\rm F}-U)\fr{m(\eps_{\rm
F}-U)}{4\pi\hbar^2} \ .
 \ee
Combining Eqs.~(\ref{psipsi_supp}) and (\ref{nr_density_supp}) we
obtain
 \bq\label{density_current_supp}
\vec{j}(\vec r) =\fr{e}{2h}\,\, \theta(\eps_{\rm F} -U(\vec
r))\,\, \vec n_z \times\vec{\nabla}U(\vec r)  \ ,
 \ee
where $\vec n_z $ is the unit vector normal to the $(x,y)$ plane.
Differentiating this with respect to $\eps_{\rm F}$ gives
$\vec{j}^{\eps_{\rm F}}(\vec r)$ of Eq.~(\ref{distrib_current}).

Our derivation of Eq.~(\ref{density_current_supp}) for a smooth
two-dimensional potential $U(\vec r)$ relies on the
nonrelativistic approximation Eq.~(\ref{Schroed_supp}). The energy
distribution of the current density $\vec{j}^{\eps}(\vec r)$ found
from Eq.~(\ref{density_current_supp}) vanishes everywhere except
in the narrow region around the line of stopping points $\eps
-U(\vec r)=0$. But the electron in a smooth potential near a
stopping point is always nonrelativistic. This suggests that the
range of validity of the result Eq.~(\ref{density_current_supp})
may be larger than just the nonrelativistic limit $|U(\vec
r)|\ll\Delta$, but it may be valid for an arbitrarily large (and
still smooth) two-dimensional potential $U(\vec r)$.

Indeed, Eqs. (\ref{current4}, \ref{currentdensity}) of the main
text (second of those is the exact analog of
Eq.~(\ref{density_current_supp})) were found for an arbitrarily
large smooth potential but only depending on one coordinate
$U(\vec r)\equiv U(x)$. That means, all what is left to do in
order to prove
Eqs.~(\ref{density_current_supp},\ref{distrib_current}) for an
arbitrarily large Fermi energy is to show that the contribution to
the anomalous current from the electrons with high energy has the
form of a local expansion in powers of the gradients of the
potential. In other words, the anomalous current at a point $\vec
r$ caused by the electrons with $\eps- U(\vec r)\gtrsim \Delta$
must be a function of the gradient, $\vec\nabla U(\vec r)$, found
at the same point.

To find the current density for an arbitrarily strong
two-dimensional potential one may take the wave function
Eq.~(\ref{psiab}) and sum up the currents due to all occupied
states. However, we may simply notice that the anomalous velocity
Eq.~(\ref{v_Magnet}) due to the solution Eq.~(\ref{psiab}) depends
only on the local derivative of the potential $U'$. This means
that if Eqs.~(\ref{distrib_current}, \ref{density_current_supp})
are valid for arbitrary $\eps_{\rm F}$ in an arbitrarily strong
uniform electric field, they should be also valid for an
arbitrarily smooth potential $U(\vec r)$ for a large Fermi energy
$\eps_{\rm F}\sim \Delta$.

\section{Appendix D: Calculation of $\beta(x)$.}

In this Appendix, we calculate explicitly starting from the
semiclassical solution of Eq.~(\ref{HamDirac}) the coefficient
$\beta(x)$ entering the wave function Eq.~(\ref{psiab}) of the
main text. This coefficient is responsible for the emergence of a
finite expectation value of the anomalous velocity $\langle v_y
\rangle$ in Eq.~(\ref{v_Magnet}). In the main text we avoid the
direct calculation of $\beta(x)$ and use instead
Eq.~(\ref{currentfinal}) to calculate $\langle v_y \rangle$.
Similar to the corresponding part of the main text, we use here
$\hbar=v_{\rm F}=1$.

Let the conduction band electron described by the massive Dirac
Hamiltonian Eq.~(\ref{HamDirac}) have an energy $\eps$ well above
the smooth potential $U(x)$ and a vanishing (conserved)
$y$-momentum, $p_y=0$. We may then introduce a
coordinate-dependent classical momentum in $x$-direction
 \bq
p_x=p(x)=\sqrt{(\eps -U(x))^2-\Delta^2} \ .
 \ee
The single plane wave solution may now be presented in the form
(in the case of reflection, which we do not consider here, there will
be a superposition of two such counter-propagating solutions)
 \bq\label{psiab_supp}
\psi =e^{i\int^x p(x')dx'} [a(x)\phi_+ +b(x)\phi_-] \ ,
 \ee
where
 \begin{align}
\phi_+ =\fr{1}{\sqrt{2\sqrt{p^2+\Delta^2}}}\left(\begin{array}{cc}
\sqrt{\sqrt{p^2+\Delta^2} +\Delta}\\
\sqrt{\sqrt{p^2+\Delta^2} -\Delta}
\end{array}\right) \ , \nonumber\\
\phi_- =\fr{1}{\sqrt{2\sqrt{p^2+\Delta^2}}}\left(\begin{array}{cc}
\sqrt{\sqrt{p^2+\Delta^2} -\Delta}\\
-\sqrt{\sqrt{p^2+\Delta^2}+\Delta}
\end{array}\right)  ,
 \end{align}
are the positive and negative energy eigenvectors in the limit of
a flat $U(x)$.

The wave function Eq.~(\ref{psiab_supp}) is in principle exact,
provided one can find the coefficients $a(x)$ and $b(x)$ to all
orders in the small $U', U'', \cdots$. Substituting
Eq.~(\ref{psiab_supp}) into the Dirac equation leads to an (exact)
system of linear equations
 \begin{align}\label{systemSC}
-ia' +i\fr{U'\Delta}{2p\sqrt{\Delta^2+p^2}} b +2\Delta b =0 \ , \\
-ib' -i\fr{U'\Delta}{2p\sqrt{\Delta^2+p^2}} a +2p \, b =0 \ .
\nonumber
 \end{align}

Approximate solutions of the system of equations (\ref{systemSC})
(which we are interested in) may be found iteratively. First, in
the second equation, we may neglect a small derivative $b'$
compared to $2pb$, leading to
 \bq\label{bSemicl_supp}
b\approx i\fr{U'\Delta}{4p^2\sqrt{\Delta^2+p^2}}\, a \ .
 \ee
Substituting this into the first equation of Eq.~(\ref{systemSC}) and
neglecting the small second order term $\sim U'b$ we find
 \bq\label{aSemicl_supp}
a\approx\sqrt{\fr{\sqrt{\Delta^2 +p^2}}{p}} \ .
 \ee
The fact that this solution reproduces correctly the classical
electron density
 \bq
\rho\approx |a|^2\sim 1/v_x \ ,
 \ee
is an additional crosscheck.

Eq.~(\ref{psiab}) of the main text is reproduced after we notice that
 \bq
\beta = ib/a \ .
 \ee

\section{Appendix E: Magnetization current {\it vs.} transport AHE
current.}

In this Appendix, we consider in more details the division of the
total anomalous Hall current into the transport and magnetization
currents discussed in the last part of the main paper. Like in the
main text, we put $\hbar = v_{\rm F}=1$. Also like in the main
text we are going to consider the transverse shift and transverse
current distribution in a wide electron ray injected parallel to
the $x$-axis and parallel to the electric field ($U(\vec r)\equiv
U(x)$) described by the wave function Eq.~(\ref{ray1}). Ray
injection with an arbitrary incident angle may be considered with
the help of Eq.~(\ref{finitepy}) of the main text. It is
constructive to consider the envelope function $g(y)$
(Eq.~(\ref{ray2})) to be flat, $g(y)\approx 1$, over the large
region $\delta y$ (i.e. $p_x\, \delta y \gg 1$) with also very
smooth steps towards $g=0$ outside this region, cf.
Fig.~\ref{fig:3}.

By calculating the longitudinal current
$j_x=e\Psi^\dagger\sigma_x\Psi$ and the density
$\rho=e\Psi^\dagger \Psi$ from Eq.~(\ref{PsiShifted}) we find
 \begin{align}\label{CurrentDensity0_supp}
j_x=eg^2\left(y+\fr{1}{2\Delta}\right)\psi_+^\dagger\psi_+ -
eg^2\left(y-\fr{1}{2\Delta}\right)\psi_-^\dagger\psi_-  \ ,
\nonumber\\
\rho =eg^2\left(y+\fr{1}{2\Delta}\right)\psi_+^\dagger\psi_+ +
eg^2\left(y-\fr{1}{2\Delta}\right)\psi_-^\dagger\psi_- \ .
 \end{align}
We remind that $\Psi=\Psi(x,y)$ is the wave function of a ray
propagating mostly in $x$-direction, Eq.~(\ref{ray1}), and
$\psi=\psi(x)$ is the plane-wave solution Eq.~(\ref{psiab}) with
momentum parallel to the electric field ($p_y\equiv 0$). The
components of $\psi(x)$ in the $\sigma_x$ eigenvalue basis are
$\psi_+=\fr{1}{2}(1+\sigma_x)\psi$ and
$\psi_-=\fr{1}{2}(1-\sigma_x)\psi$. The second equation
(\ref{CurrentDensity0_supp}) was used in the main text to
calculate the side-jump of the trajectory Eq.~(\ref{y(x)}).

Now we may use the fact that for our choice of the normalization
of the wave function Eq.~(\ref{psiab}) $\psi_+^\dagger\psi_+
-\psi^\dagger_-\psi_-=1$ and $\psi_+^\dagger\psi_+
+\psi^\dagger_-\psi_-=1/v_x$ and write instead of
Eq.~(\ref{CurrentDensity0_supp})
 \begin{align}\label{CurrentDensity_supp}
j_x&=eg^2\left(y\right) + eg'g/(\Delta v_x) \ ,
\nonumber\\
\rho &=eg^2\left(y\right)/v_x + eg'g/\Delta \ ,
 \end{align}
where $v_x=v_x(x)$ acquires a dependance on $x$ in the case of a
smooth potential $U(x)$,
 \bq
v_x=\fr{p(x)}{\sqrt{\Delta^2+p(x)^2}} \ , \
p(x)=\sqrt{(\eps-U(x))^2-\Delta^2} \ .
 \ee

The second equality in Eq.~(\ref{CurrentDensity_supp}) may be
rewritten as
 \bq
\rho =eg^2(y+ v_x/(2\Delta))/v_x \ ,
 \ee
which immediately gives the trajectory (equivalent to
Eq.~(\ref{y(x)}) of the main text)
 \bq
y(x)=-\fr{v_x(x)}{2\Delta} \ .
 \ee
Differentiating this $y(x)$ allowed us in the main text to find
the transverse velocity $v_{y{\rm tr}}$ Eq.~(\ref{v_y}). Since
this velocity originates from the displacement of the ray and not
from the wave-packet's internal dynamics, it leads to the
transport anomalous current $j_{y{\rm tr}}=v_{y{\rm tr}}\rho$.

The possible time evolution of the quantum ray Eq.~(\ref{ray1})
should proceed in agreement with the continuity equation. Since
the ray is built from waves of the same energy, the charge
distribution is stationary and the continuity equation reduces to
the vanishing of the divergence of the current, ${\rm div}\, \vec
j=0$. Thus we write
 \bq\label{continuity AppE}
\fr{\partial j_x}{\partial x}=-e\fr{g'gv_x'}{\Delta v_x^2} =
-\fr{\partial j_y}{\partial y} \ ,
 \ee
where $g'=dg/dy$ and $v_x'=dv_x/dx$. The current $j_y$ here can
not be derived directly from the wave function Eq.~(\ref{ray2}).
It appears due to the correction $\sim \beta\sigma_y$ to the
semiclassical wave-function in Eq.~(\ref{psiab}), which was
omitted in Eq.~(\ref{ray2}) (and consequently in
Eqs.~(\ref{CurrentDensity0_supp})). Still the continuity equation
Eq.~(\ref{continuity AppE}) allows us to find this current. The
total anomalous current density $j_y$ is expected to be
independent of the transverse coordinate $y$ only inside the ray,
where $g(y)\approx 1$. At the borders of the ray $j_y$ increases
from zero (outside) to this constant value. This increase is
described by Eq.~(\ref{continuity AppE}). Integration over $y$ of
the second equality in Eq.~(\ref{continuity AppE}) gives the value
of the anomalous current inside the ray
 \bq
j_y=e\fr{g^2v_x'}{2\Delta v_x^2} \ .
 \ee
The calculation of the derivative of the longitudinal velocity
$v_x'$ here gives
 \begin{align}
v_x'=\fr{dv_x}{dp}\fr{dp}{dx}&=\fr{\Delta^2}{(\Delta^2+p^2)^{3/2}}
\times \fr{-U'}{p}\sqrt{\Delta^2+p^2} \nonumber\\
&=
\fr{-U'\Delta^2}{p(\Delta^2+p^2)} \ ,
 \end{align}
leading to
 \begin{align}\label{jyvy supp}
j_y&=-eg^2U'\fr{\Delta}{2p^3} \ , \\
\langle v_y\rangle
&=\fr{j_y}{\rho}=\fr{-U'\Delta}{2p^2\sqrt{\Delta^2+p^2}} \
,\nonumber
 \end{align}
in agreement with Eq.~(\ref{v_Magnet}) of the main text. Thus we
see that the anomalous transverse current inside the ray (and the
velocity Eq.~(\ref{v_Magnet})), which is not captured by
Eqs.~(\ref{ray1}, \ref{ray2}, \ref{PsiShifted}) follows from them
through the continuity equation. Eq.~(\ref{jyvy supp}) shows the
total microscopic AHE current for which the corresponding velocity
$\langle v_y\rangle$ can not be deduced from Eq.~(\ref{Berry}).

As was written in the main text, the difference between the total
microscopic and transport current densities is naturally
attributed to the magnetization current~\cite{CooperPRB97}
 \bq
\vec j=\vec j_{tr} +\vec j_{mag} \ , \ \vec j_{mag} =\vec\nabla
\times \vec M(\vec r ) \ ,
 \ee
with $\vec M(\vec r )$ being the density of the magnetic moment.
According to Ref.~\cite{NiuPRL07} the magnetic moment of an
electron subject to the massive two-dimensional Dirac Hamiltonian
is
 \bq
\mu(p)=\fr{e}{2\hbar}\fr{m v_{\rm F}^2}{m^2v_{\rm
F}^2+p^2}=\fr{e\Delta}{2(\Delta^2+p^2)} \ .
 \ee
That gives the magnetization density
 \bq
M=\mu\Psi^\dagger\Psi =\mu/v_x
=\fr{e\Delta}{2p\sqrt{\Delta^2+p^2}} \ .
 \ee
Consequently, we find, in agreement with the general
expectation~\cite{CooperPRB97,XiaoPRL06},
 \bq
-\fr{\partial M}{\partial x}= -\fr{eU'\Delta}{2}\left( \fr{1}{p^3}
+\fr{1}{p(\Delta^2+p^2)}\right) = j_y -j_{y_{tr}} \ .
 \ee
With that we have explicitly demonstrated that the spatially
inhomogeneous magnetization is responsible for the difference
between the total current and the transport current.

\end{document}